\documentclass[twocolumn,trackchanges]{aastex62}

\usepackage{rotating}

\graphicspath{{./}{figures/}}

\submitjournal{ApJ}

\shorttitle{Bootes III is disrupting}
\shortauthors{Carlin \& Sand}

\begin{document}

\title{Bo\"{o}tes~III is a disrupting dwarf galaxy associated with the Styx stellar stream}

\correspondingauthor{Jeffrey L. Carlin}
\email{jcarlin@lsst.org}

\author[0000-0002-3936-9628]{Jeffrey L. Carlin}
\affil{LSST, 950 North Cherry Avenue, Tucson, AZ 85719, USA} 

\author{D. J. Sand}
\affil{Department of Astronomy/Steward Observatory, 933 North Cherry Avenue, Rm. N204, Tucson, AZ 85721-0065, USA}

\begin{abstract}

We present proper motion measurements of Bo\"{o}tes~III, an enigmatic stellar satellite of the Milky Way, utilizing data from the second data release of the {\it Gaia} mission.  By selecting 15 radial velocity confirmed members of Bo\"{o}tes~III, along with a likely RR Lyrae member in the vicinity, we measure an error weighted mean proper motion of $(\mu_{\alpha} \cos{\delta}, \mu_\delta) = (-1.14, -0.98)\pm(0.18, 0.20)$~mas~yr$^{-1}$.  We select and present further stars that may be Bo\"{o}tes~III members based on their combined proper motion and position in the color magnitude diagram. We caution against assigning membership to stars that are not confirmed spectroscopically, as we demonstrate that there are contaminating stars from the disrupting globular cluster NGC~5466 in the vicinity of the main body of Bo\"{o}tes~III, but we note that our results are consistent with previous Bo\"{o}tes~III proper motion estimates that did not include spectroscopic members.  Based on the measured proper motion and other known properties of Bo\"{o}tes~III, we derive its Galactocentric velocity and compute its orbit given canonical Milky Way potentials with halo masses of both 0.8$\times$10$^{12}$ M$_{\odot}$ and  1.6$\times$10$^{12}$ M$_{\odot}$.  These orbits robustly show that Bo\"{o}tes~III passed within $\sim$12 kpc of the Galactic center on an eccentric orbit roughly $\sim$140 Myr ago.  Additionally, the proper motion of Bo\"{o}tes~III is in excellent agreement with predictions for the retrograde motion of the coincident Styx stellar stream.  Given this, along with the small pericenter and metallicity spread of Bo\"{o}tes~III itself, we suggest that it is a disrupting dwarf galaxy giving rise to the Styx stellar stream.
\end{abstract}



\section{Introduction} \label{sec:intro}

The orbits of the Milky Way (MW) satellites  provide another dimension in understanding their formation and evolution.  This has been most clearly shown for the Magellanic Clouds, whose proper motions \citep{Kall06,Kall13} suggest that they are on their first passage through the Milky Way \citep{Besla07}, changing our historical view of the Magellanic Stream \citep[for a recent review, see][]{D16}.  Until recently, proper motion measurements for distant MW satellites have required long baselines and (often) the precision of the {\it Hubble Space Telescope} \citep[e.g.,][among others]{Sohn13,Pryor15,Dana16,Piatek16,Sohn17,Fritz17,Dana18}.

The second data release (DR2) of the {\it Gaia} mission \citep{GaiaDR2} included proper motion data with typical uncertainties of $\approx$1.2 mas yr$^{-1}$ for stars with $G$$\approx$20 mag, allowing proper motions and orbits for much of the MW's satellite population to be calculated \citep{Helmi18,Simon2018,Fritz18,Kall18,Watkins18,Massari18}.  Among the ultra-faint dwarf galaxies, only a handful of systems (e.g., Tucana~III, Crater~II, and Segue~2; \citealt{Simon2018,Fritz18}) appear to be on orbits whose pericenters are near enough to the MW for them to have experienced significant tidal disturbance, in contrast to findings of some previous imaging and spectroscopic studies \citep[e.g.,][among others]{Sand09,Munoz10,Sand12,Roderick15,Collins17,Garling18}, although the orbits of several systems are still too uncertain to draw definitive conclusions.    

Here we focus on the proper motion and orbit of the enigmatic system Bo\"otes~III (BooIII).  BooIII was discovered as a stellar overdensity, spanning $\sim$1~deg on the sky, nearly coincident with the Styx stellar stream \citep{Grillmair2009}.   At the inferred distance of $\approx$47 kpc, BooIII is roughly 1 kpc in size (analogous to the diffuse system Crater II; \citealt{Torrealba16}).  Its main body has an absolute magnitude of $M_{\rm V}\approx-5.8$ mag  \citep{Correnti2009}.  \citet{Carlin2009} identified 20 candidate BooIII member stars via spectroscopic follow-up, and derived a systemic $V_{\odot}$=197.5$\pm$3.8 km s$^{-1}$, and velocity dispersion of $\sigma$=14.7$\pm$3.7 km s$^{-1}$, possibly inflated due to its dynamical state. This same spectroscopic sample indicated a stellar population with a significant spread in metallicity. Because of its distended morphology, metallicity spread, and possible association with the Styx stellar stream, BooIII has long been a candidate disrupting dwarf system \citep{Grillmair2009,Correnti2009,Carlin2009}, and now with the advent of {\it Gaia} DR2, this can be critically assessed via its inferred orbit.  In \S 2 we gather the relevant {\it Gaia} DR2 data and  measure the proper motion of BooIII.  We also select and present further potential BooIII member stars based on their position on the color magnitude diagram and their proper motion.  In \S~3 we calculate the orbit of BooIII based on standard parameterizations of the Milky Way potential, and discuss BooIII's relationship with the Styx stellar stream.  We summarize and conclude in \S~4.

\section{Data and Analysis} \label{sec:data}

We present the absolute proper motion (PM) of BooIII. Members of BooIII were selected based on spectroscopic velocities from \citet{Carlin2009}, who found a systemic velocity of $V_{\odot} = 197.5$~km~s$^{-1}$ from 20 member stars. These members include 6 blue horizontal branch (BHB) stars at a magnitude of $g_0\sim$19,\footnote{Throughout this work, we use colors and magnitudes from the PanSTARRS-1 (PS1) survey \citep{Magnier2016,Flewelling2016,Chambers2016_PS1}. All magnitudes are corrected for extinction using $E(B-V)$ values for each star from the maps of \citet{Schlegel1998}, with coefficients for the PS1 bands from  \citet{Schlafly2011}.} while all of the other members are fainter, lower-RGB stars.

\begin{figure}[!t]
\includegraphics[width=1.0\columnwidth]{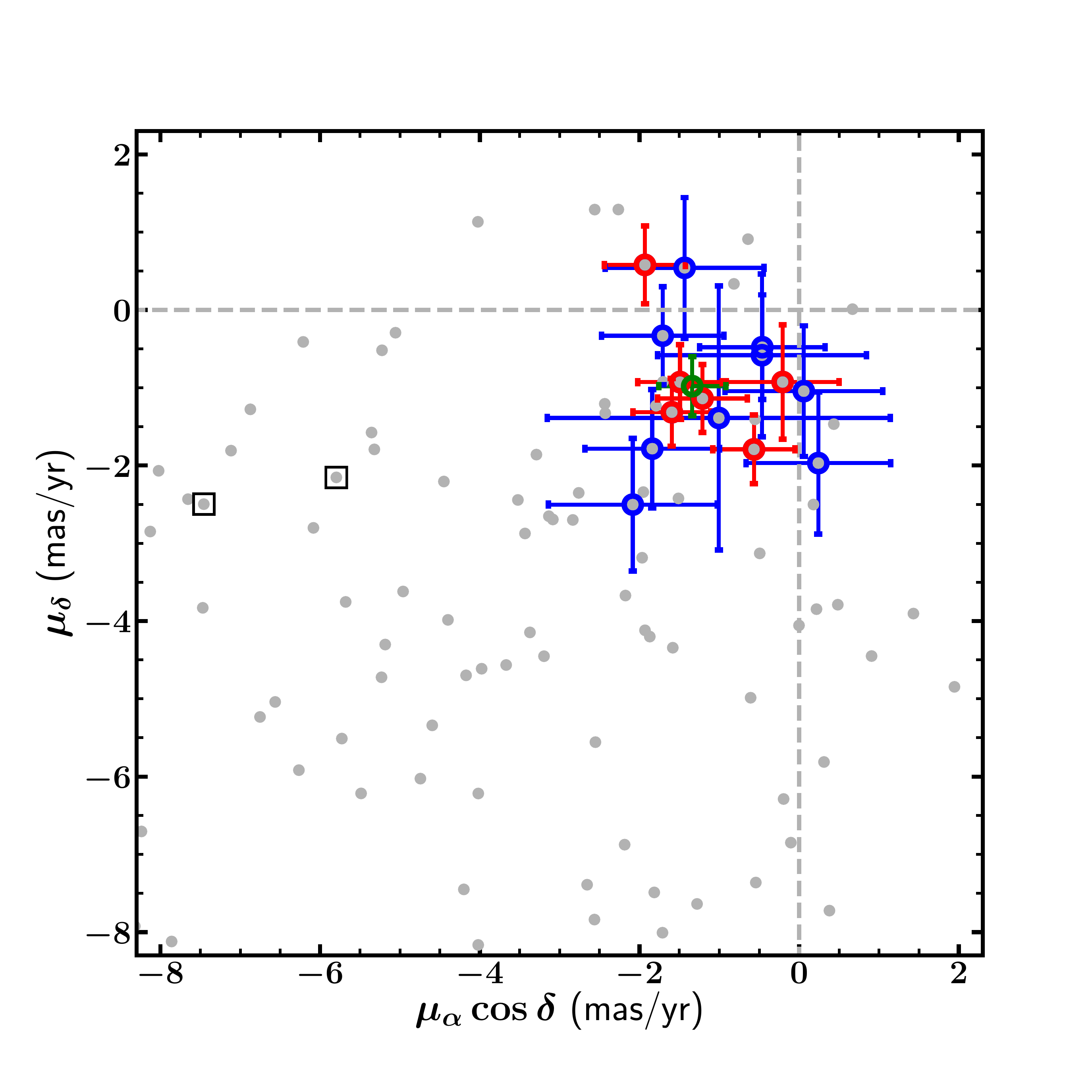}
\caption{Proper motions of stars from the radial velocity sample of \citet{Carlin2009}. Red points are RV-confirmed BooIII BHB members, blue points are lower-RGB RV members, and the green point is the known RR Lyrae star from \citet{Sesar2014}. Open squares are RV candidates from \citet{Carlin2009} whose membership in BooIII is ruled out by their large proper motions. Note that two additional non-members have proper motions outside the boundaries of this figure.}
\label{fig:pms}
\end{figure}

\subsection{Proper motion of Bo\"otes III}

Of the 194 stars from the \citet{Carlin2009} spectroscopic sample, 174 have matches in the Gaia DR2 catalog, including 19 of the 20 candidate radial velocity (RV) members. All matches from our nearest-neighbor search have positional offsets of $<0.3"$. We confirmed that the matches are reasonable by verifying that our original SDSS magnitudes match those from PS1 within 0.1 mag (for all but 6 stars; all stars agree within 0.25 mags). 
The proper motions of 17 of these stars are shown in Figure~\ref{fig:pms} (the other two have large proper motions that place them outside the boundaries of Figure~\ref{fig:pms}). For comparison, we include all other stars from the \citet{Carlin2009} sample as gray points. BooIII velocity candidates (red and blue points with error bars, and open black squares) mostly clump together; we remove the two stars shown as open squares, plus two more beyond the plot boundaries, because they have proper motions that are too large for Milky Way stars at $\sim50$~kpc. This leaves a final sample of 15 BooIII RV members; 6 BHB members are shown as red symbols in Fig.~\ref{fig:pms}, and the other 9 RGB candidates are blue symbols. 

\citet{Sesar2014} identified an RR Lyrae star from the Palomar Transient Factory \citep[PTF; ][]{Law2009PS1,Rau2009PS1} data that is at a distance and radial velocity consistent with being associated with BooIII. This star is $\sim1^\circ$ southeast of the nominal center of BooIII derived by \citet{Grillmair2009}. We checked the proper motion of this star, and it is also consistent with BooIII (see green symbol in Figure~\ref{fig:pms}), and we include it in our list of members, bringing the total to 16 members. The properties of these 16 stars, plus the five proper motion outliers we removed, are given in Table~\ref{tab:members}.

The 6 BHB stars have mean proper motion $(\mu_{\alpha} \cos{\delta}, \mu_\delta) = (-1.17, -0.92)\pm(0.24, 0.30)$~mas~yr$^{-1}$, while the error-weighted mean proper motion of all 16 member stars is $(\mu_{\alpha} \cos{\delta}, \mu_\delta) = (-1.14, -0.98)\pm(0.18, 0.20)$~mas~yr$^{-1}$. Because these results are consistent within the uncertainties, we will use the proper motion from all 16 stars for subsequent analysis.

\begin{figure}[!t]
\includegraphics[width=0.95\columnwidth]{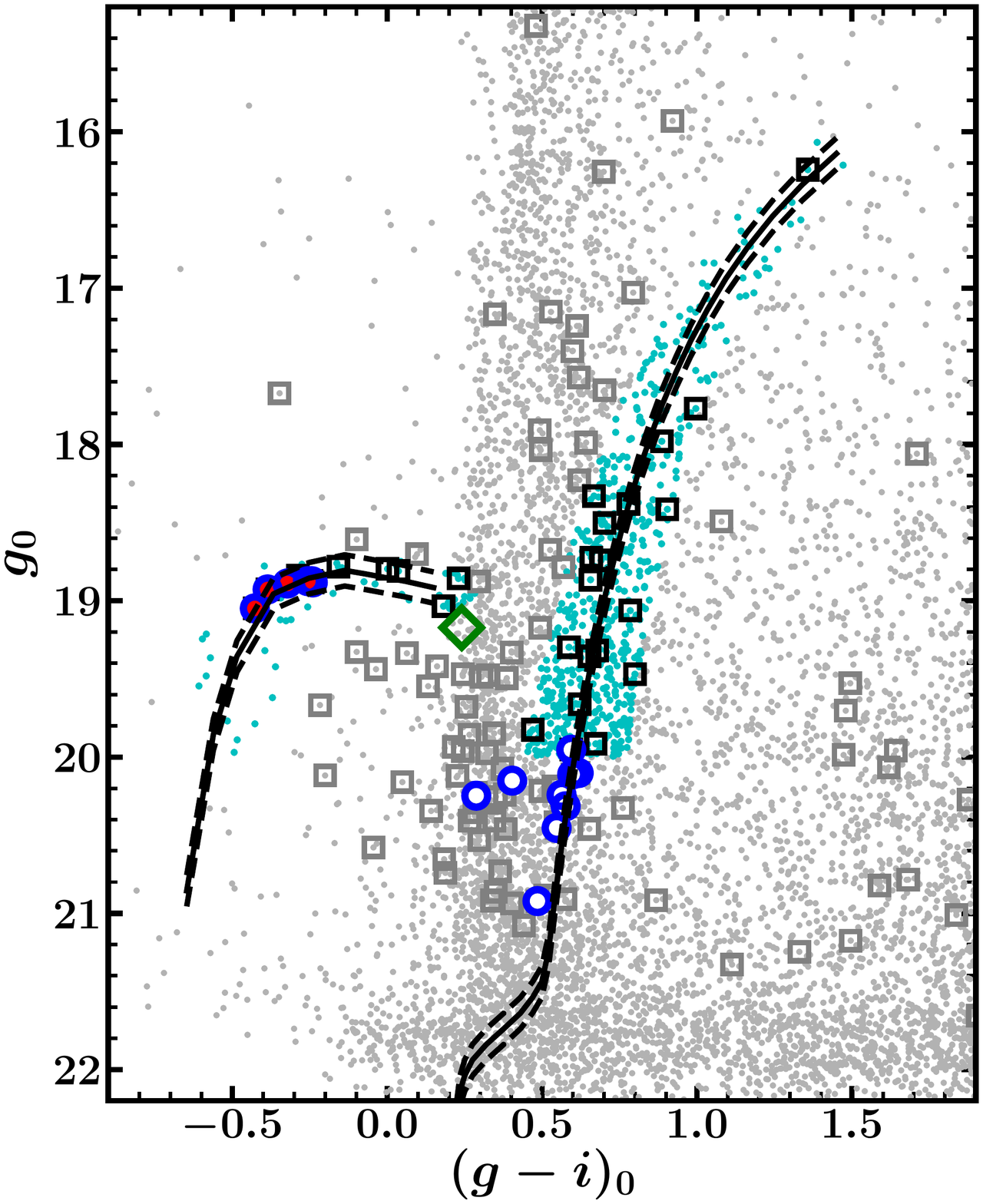}
\caption{PS1 CMD of all stars within $1^\circ$ of BooIII (small gray points). Red circles are RV-confirmed BHB members, blue points are lower-RGB RV members, and the green diamond is the RR Lyrae star from \citet{Sesar2014}. The empirical ridgeline from \citet{Bernard2014} for globular cluster M~15 ([Fe/H]=-2.34), shifted to a distance of 46.5 kpc, is shown as a solid black line, with dashed isochrones on either side shifted by $\pm2.0$ kpc. Large open squares are stars with proper motions between -3.0 to 1.0~mas~yr$^{-1}$ in both $\mu_{\alpha} \cos{\delta}$ and $\mu_{\delta}$, with black squares showing candidate BooIII members satisfying this proper motion cut, within $30'$ of the BooIII center, and also within the M~15 CMD filter highlighted by cyan points. }
\label{fig:cmd_ps1}
\end{figure}

\subsection{Refining the distance to BooIII}

We re-derive the distance to BooIII with data from PS1. Specifically, we use the empirical ridgeline of metal-poor ([Fe/H] = -2.34; \citealt{Carretta09}) globular cluster M~15 (NGC~7078) derived by \citet{Bernard2014} from PS1 photometry. A least-squares fit of the M~15 horizontal branch to the BooIII BHB stars yields a BooIII distance modulus of $(m-M)_0 = 18.34\pm0.02$ mag (assuming $m-M = 15.39$ for M~15). This agrees well with the BooIII measurement by \citet{Grillmair2009} of $18.35\pm0.01$ mag. The formal uncertainty on our measured distance modulus would correspond to an uncertainty of $\sim0.4$~kpc. However, given the small number of stars used to constrain the fit, and possible uncertainties in the distance to M~15, we conservatively adopt a BooIII distance of $d_{\rm BooIII} = 46.5\pm2.0$~kpc.

\subsection{A search for further BooIII member stars}
The 15 RV members of BooIII are highlighted in the color magnitude diagram (CMD) presented in Fig.~\ref{fig:cmd_ps1} as large blue circles, with the BHB stars filled in as red points. We additionally include (as a green open diamond) the RR Lyrae star from \citet{Sesar2014} that is $\sim1^\circ$ from our adopted center of BooIII, with a distance and radial velocity consistent with BooIII membership. The background CMD shows all stars from PS1 \citep{Chambers2016_PS1} within $1^\circ$ of the BooIII center as gray points. The PS1 ridgeline of M~15 is shown as a solid black line in the CMD (Figure~\ref{fig:cmd_ps1}), with dashed lines on either side illustrating a $\pm2.0$~kpc distance range about the mean of 46.5~kpc. We created a CMD filter based on this ridgeline, with width of 0.1~mag at $g_0=16$, increasing linearly to 0.2~mag width at $g_0=22.5$; stars selected with this filter, and at $g_0<20$, are highlighted as cyan points in Fig.~\ref{fig:cmd_ps1}.\footnote{Given the possible metallicity spread of BooIII \citep{Carlin2009}, member stars are expected to have large scatter about the RGB shown in Figure~\ref{fig:cmd_ps1}. We chose a fairly narrow filter in order to conservatively identify a more secure sample of members, and avoid contamination from, e.g., NGC 5466 stars in the vicinity, at the possible expense of sacrificing some BooIII members.} We matched the PS1 catalog to {\it Gaia} DR2, and selected stars with proper motions consistent with our measurement for BooIII, in hopes of isolating bright RGB stars with precise {\it Gaia} proper motions. These proper-motion selected stars, with $-3 < \mu_{\alpha} \cos{\delta} < 1$~mas~yr$^{-1}$ and $-3 < \mu_{\delta} < 1$~mas~yr$^{-1}$, within $30'$ of the BooIII center and also selected within the CMD filter, are shown as open black squares (PM-selected stars that are outside the CMD filter are gray squares), and their coordinates, magnitudes, and proper motions are given in Table~\ref{tab:pm_cands}. There are only a handful of stars along the RGB ridgeline with an appropriate proper motion (and many of these may be contaminants; see below), suggesting that BooIII, like many other ultra-faint dwarfs \citep[see][for deep CMDs of comparable luminosity systems]{Sand12,Carlin17,Mutlu18}, has few RGB stars. 

The error-weighted average proper motions of the 23 stars shown as open squares in Fig.~\ref{fig:cmd_ps1} and at $g_0 < 20$ are $(\mu_{\alpha} \cos{\delta}, \mu_\delta) = (-1.04, -1.02)\pm(0.17, 0.16)$~mas~yr$^{-1}$. These are consistent (within the uncertainties) with the proper motions of the RV members. This is perhaps not surprising, because we selected them within a strict proper motion box centered on these values. However, we note that it is particularly important to use confirmed radial velocity members in order to accurately determine the proper motion of BooIII. This is necessary in order to mitigate contamination from the nearby, disrupting globular cluster NGC~5466. Stars from NGC~5466 are clearly visible among the PM-selected candidates (gray squares) in Fig.~\ref{fig:cmd_ps1}, forming a main sequence turnoff at $g_0 \sim 19.8$ (i.e., consistent with the $\sim16$~kpc distance to NGC 5466). NGC 5466 is $\sim2.5^\circ$ from BooIII at a $V_{\odot}$=111 km s$^{-1}$, has a half-light radius of $r_{\rm half} \sim 2.3'$ \citep{Harriscat}, and a proper motion of $(\mu_{\alpha} \cos{\delta}, \mu_\delta) = (-5.40, -0.79)\pm(0.004, 0.004)$~mas~yr$^{-1}$ \citep{Helmi2018}.  Deep imaging has shown that NGC~5466 is disrupting, with a prominent tidal stream \citep{Grillmair06,Belokurov06},  so it is not surprising to find members of this globular cluster in our 30 arcmin selection radius. 

Since the stars that appear in our simple proper motion (and, to some extent, isochrone-filtered) selection seem to be contaminated by NGC~5466 stars, this suggests that selecting BooIII members only based on generous CMD and PM selections as in \citet{Massari18} is likely to bias the resulting mean proper motion. We further note that globular cluster NGC~5272 (M~3) is $\sim3.5^\circ$ from BooIII at a distance of $\sim10.2$~kpc, and has a proper motion within the box selection we used above. Thus, it is also possible that this cluster could contaminate samples that do not rely on confirmed RV members. However, given the agreement between our measured proper motions and those of \citet{Massari18}, it appears that contamination from the nearby globular clusters (and possibly associated streams) has only a small effect on the results.

\begin{table*}
\begin{center}
\caption{Kinematic and orbital parameters of Bo\"otes III.}
\begin{tabular}{lccc}
\hline
\hline
parameter & BHB stars only & all members,\tablenotemark{a} & all members,\\
 & low-mass MW & low-mass MW & massive MW\\
\hline
N$_{\rm stars}$ & 6 & 16 & 16 \\
distance (kpc) & $46.5\pm2.0$ & $46.5\pm2.0$ & $46.5\pm2.0$\\
$V_{\odot}$ (km s$^{-1}$)\tablenotemark{b} & $197.5\pm3.8$ & $197.5\pm3.8$ & $197.5\pm3.8$\\
$\mu_{\alpha} \cos{\delta}$ (mas yr$^{-1}$) & $-1.17\pm0.24$  & $-1.14\pm0.18$ & $-1.14\pm0.18$\\
$\mu_{\delta}$ (mas yr$^{-1}$) & $-0.92\pm0.30$ & $-0.98\pm0.20$ & $-0.98\pm0.20$\\
$(U, V, W)$ (km s$^{-1}$) & (-17.0, -288.3, 251.5) & (-3.6, -294.2, 249.5) & (-3.6, -294.2, 249.5)\\
$r_{\rm peri}$ (kpc) & $12.4^{+6.1}_{-6.1}$ & $12.6^{+5.6}_{-5.2}$ & $10.6^{+5.4}_{-4.5}$\\
$r_{\rm apo}$ (kpc) & $196.6^{+102.1}_{-48.4}$ & $193.0^{+76.2}_{-41.5}$ & $88.9^{+14.0}_{-8.4}$\\
eccentricity & $0.89^{+0.03}_{-0.03}$ & $0.88^{+0.03}_{-0.02}$ & $0.79^{+0.07}_{-0.06}$\\
period (Gyr) & $3.5^{+2.5}_{-1.1}$ & $3.5^{+1.8}_{-0.9}$ & $1.0^{+0.2}_{-0.1}$\\
\hline \\
\end{tabular} 
\tablenotetext{a}{Includes 15 RV members plus known RR Lyrae star from \citet{Sesar2014}.}
\tablenotetext{b}{From \citet{Carlin2009}.}
\label{tab:orb_params}
\end{center}
\end{table*}

\begin{figure*}[!t]
\includegraphics[width=0.5\textwidth]{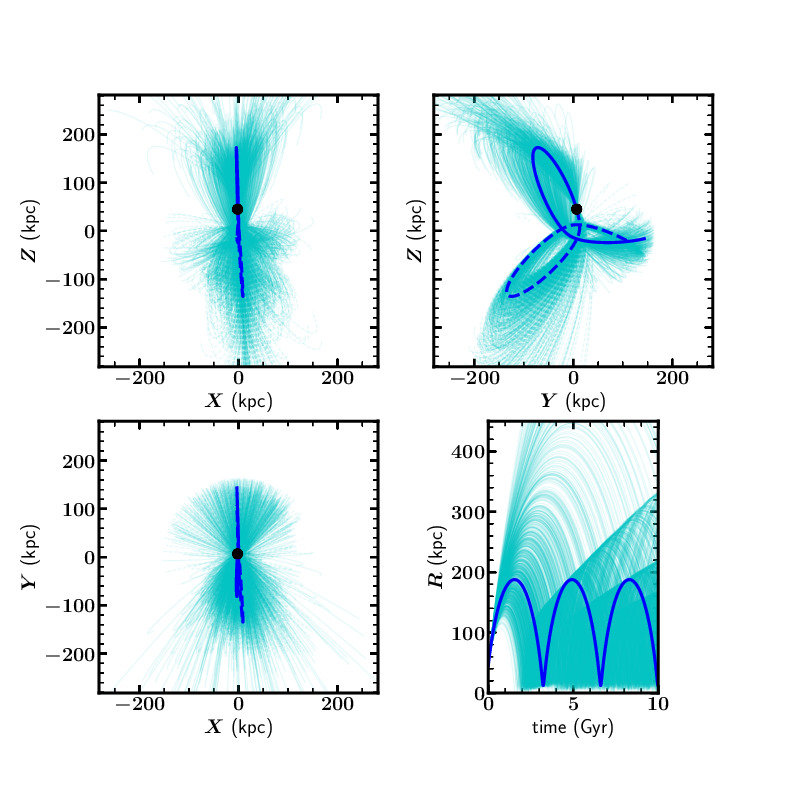}
\includegraphics[width=0.5\textwidth]{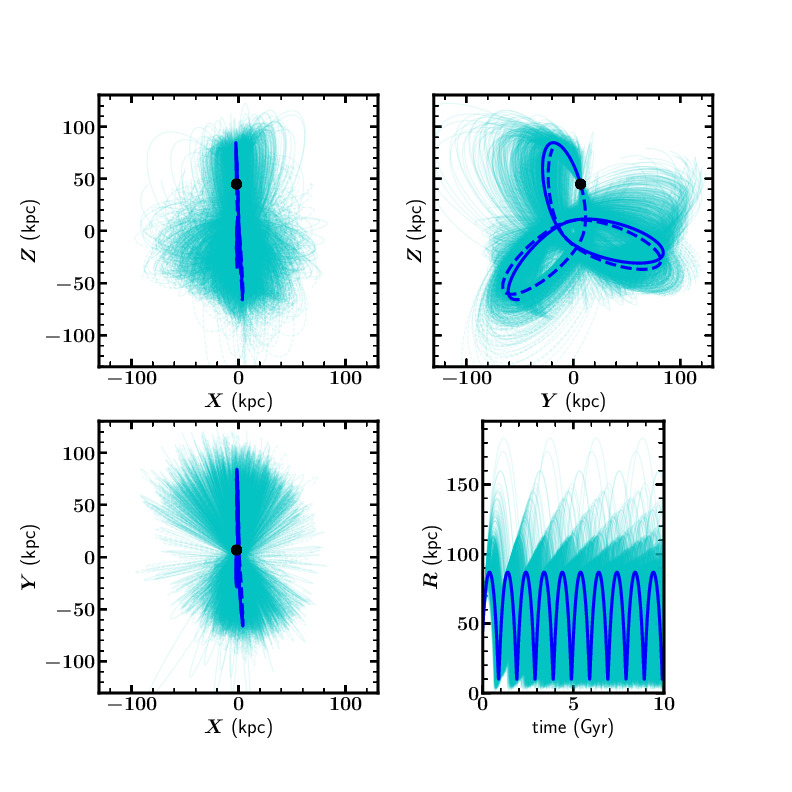}
\caption{Projections of the orbit of BooIII in Galactic Cartesian coordinates, as well as its MW distance as a function of time. The light cyan orbits represent different realizations drawn from the error distributions on our measured distances, velocities, and proper motions. Dark blue lines represent the orbit based on the mean properties of BooIII, with solid lines showing the forward integration, and dashed lines the past trajectory of BooIII. {\it Left:} MW halo mass 0.8$\times$12 M$_{\odot}$ and a 4 Gyr orbit integration; {\it Right:} MW halo mass of 1.6$\times$12 M$_{\odot}$, and a 2.5 Gyr integration.}
\label{fig:orbit}
\end{figure*}

\section{The Orbit of Bo\"otes III}

With kinematics in hand, we now explore the orbit of BooIII. Given its extended nature and distorted morphology \citep{Grillmair2009}, as well as its large velocity dispersion and metallicity spread \citep{Carlin2009}, we are particularly interested in testing whether BooIII is on an orbit that would have recently subjected it to significant tidal influence. We integrate orbits in a model Galactic potential using the \texttt{galpy} package \citep{Bovy2015galpy}. In particular, we use the ``default'' \texttt{MWPotential2014} model Galactic potential, which consists of a Miyamoto-Nagai disk, spherical bulge with power-law density profile, and an NFW \citep{NFW1997} halo (see parameters in \citealt{Bovy2015galpy}). All calculations in this work assume a Solar radius of $R_0 = 8.0$~kpc, circular velocity at $R_0$ of $V_0 = 220$~km~s$^{-1}$, and the \citet{Schoenrich2010} Solar motion. Note that, as in \texttt{galpy}, we adopt left-handed Galactic Cartesian coordinates (velocities) with $X (U)$ positive in the direction toward the Galactic center, $Y (V)$ oriented along Galactic rotation, and $Z (W)$ toward the north Galactic pole.

The resulting orbit based on the mean values of measured parameters from Table~\ref{tab:orb_params} is shown as the dark blue curve in the left portion of Figure~\ref{fig:orbit}. The motion is confined mostly to the Galactic $Y-Z$ plane, taking BooIII to peri- and apo-center distances of $\sim12$ and $\sim193$~kpc, respectively -- i.e., a rather eccentric orbit ($e = 0.88$). To place error bars on the orbital parameters, we integrate 1000 orbits, randomly selecting values from Gaussian distributions centered on the mean values of distance, $V_{\odot}$, and proper motions, with $\sigma$ equal to the uncertainties quoted in Table~\ref{tab:orb_params}. We show these 1000 orbits as faint cyan-colored paths in Fig.~\ref{fig:orbit}. The mean BooIII orbital parameters and their uncertainties are estimated as the 50th (median), 14th, and 86th percentiles of the distributions resulting from the 1000 integrations, and given in Table~\ref{tab:orb_params}. A small fraction of the orbits result in BooIII being unbound from the Milky Way. 

The NFW halo implemented in \texttt{galpy} has a low mass relative to recent measurements of the MW halo potential \citep[e.g.,][see summary of the wide range of MW mass estimates in \citealt{BlandHawthornGerhard2016}]{Watkins18}, which find a halo mass closer to $\sim1.5\times10^{12}~M_\odot$, or roughly twice that of the \texttt{galpy} halo (which has $0.8\times10^{12}~M_\odot$). We thus explore orbits in a modified version of the \texttt{MWPotential2014} with the halo mass increased to $1.6\times10^{12}~M_\odot$. These are shown in the right half of Figure~\ref{fig:orbit}, with the derived orbital parameters given in the ``massive MW'' column of Table~\ref{tab:orb_params}. The orbit in this more massive potential is still rather eccentric, with a similar pericenter as the low-mass MW model, but reaches an apocenter of only $\sim90$~kpc, with a much shorter period of $\sim1$~Gyr. Clearly the halo potential has a dramatic effect on radial orbits such as that of BooIII, which in turn suggests that BooIII may provide a sensitive probe to constrain the MW halo mass. 

\begin{figure}[!t]
\includegraphics[width=0.95\columnwidth]{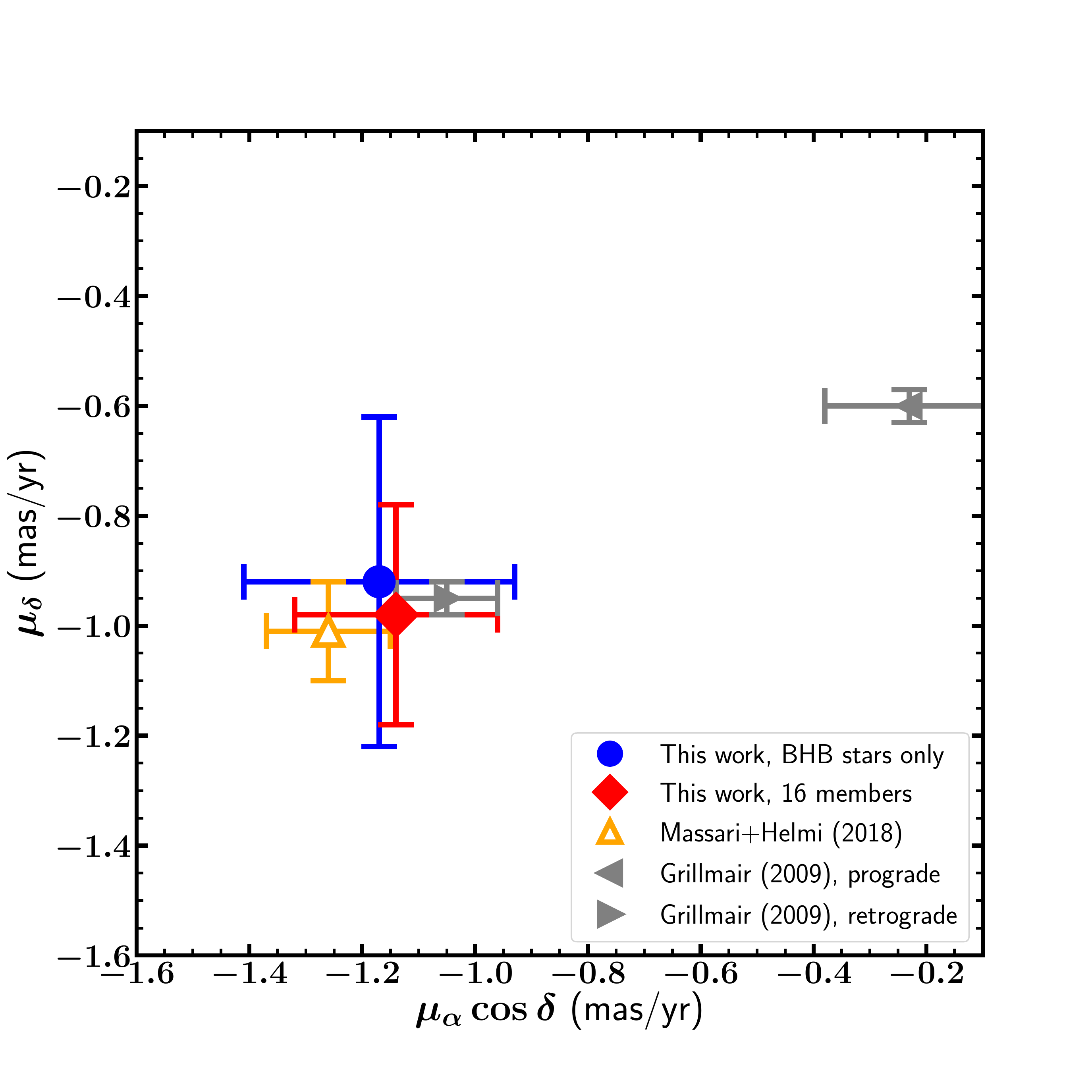}
\caption{Measured proper motions of BooIII. Our estimate based on 6 BHB members is shown in blue, and the measurement from 15 RV members plus one known RR Lyrae star is the red diamond. The measurement from \citet{Massari18} is an orange triangle. For comparison, we show the predicted proper motions based on orbit derivations from \citet{Grillmair2009}. Our measurements clearly rule out the prograde orbit, and are consistent with the retrograde Styx stream orbit.}
\label{fig:compare_pms}
\end{figure}

\subsection{Are BooIII and the Styx stream related?}

In the discovery paper announcing BooIII \citep{Grillmair2009}, it was noted that the Styx stream (also first seen in the same work) apparently passes through the position of BooIII, and is at roughly the same distance as BooIII, suggesting an association between them. While there are no confirmed kinematic members of the Styx stream, its relative distance from the Sun is reasonably well determined along its large angular extent on the sky. Thus \citet{Grillmair2009} was able to estimate the orbit of Styx, with the caveat that without velocities, we cannot know whether it orbits in a prograde or retrograde sense. \citet{Grillmair2009} predicted a range of radial velocities and proper motions for prograde and retrograde Styx orbits. In Figure~\ref{fig:compare_pms} we compare our measured proper motions for BooIII with the predictions by \citet{Grillmair2009}. The proper motions of BooIII are inconsistent with the prograde orbit for Styx, and agree with the predicted retrograde orbit. In order for the Styx orbit to match that of BooIII, the radial velocity of Styx at the fiducial point used by \citet{Grillmair2009} must be at the high end of their quoted range. If the Styx stream does indeed originate from BooIII (or from a common progenitor), then our derived orbit provides clear predictions for the expected kinematics and 3D spatial distribution of Styx stream stars. In particular, the orbit of BooIII suggests that the portions of Styx eastward of BooIII should be nearer to the Sun than at the fiducial point near BooIII, reaching distances of $\sim19$~kpc at $RA\sim260^\circ$ (roughly the eastern edge of the SDSS footprint). This is contrary to the findings of \citet{Grillmair2009}, who estimated that Styx stars are more distant in the eastern portion compared to the western region near BooIII.

Finally, we note that we could discern no obvious overdensity among {\it Gaia} DR2 proper motions within $5^\circ$ of BooIII that would suggest detection of the Styx stream's kinematics. This may be simply because the surface density of Styx stars is much too low to manifest as ``clumping'' in a proper motion vector point diagram such as Figure~\ref{fig:pms}. Definitively understanding the nature of Styx will likely require first identifying radial velocity members, which can then be used to pick members from {\it Gaia} with which to derive an orbit.


\begin{table}
\begin{center}
\caption{Updated properties of Bo\"otes III.}
\begin{tabular}{lc}
\hline
\hline
parameter & all members\tablenotemark{a}\\
\hline
N$_{\rm stars}$ & 16 \\
distance (kpc) & $46.5\pm2.0$ \\
$V_{\odot}$ (km s$^{-1}$) & $197.1\pm3.6$ \\
$\sigma_v$ (km s$^{-1}$) & $10.7\pm3.5$ \\
${\rm [Fe/H]}$\tablenotemark{b} & $-2.1\pm0.2$\\
$\sigma_{\rm [Fe/H]}$ & $0.55\pm0.19$\\
\hline \\
\end{tabular} 
\tablenotetext{a}{Includes 15 RV members plus known RR Lyrae star from \citet{Sesar2014}.}
\tablenotetext{b}{Based on 9 RGB members and the RR Lyrae star; excludes BHB stars.}
\label{tab:meas_params}
\end{center}
\end{table}

\section{Conclusions \& Summary}\label{sec:conclude}

BooIII passed within $\sim12$~kpc of the Galactic center on its eccentric orbit only $\sim140$~Myr ago. A dwarf galaxy on such an orbit would have been subject to significant tidal forces, which readily explains the distorted morphology of BooIII \citep{Carlin2009,Grillmair2009}. Furthermore, we have shown that our derived orbit is directed along the Styx stream, confirming the association between BooIII and Styx. Because its orbit traces a large radial extent (from $12\lesssim R_{\rm GC} \lesssim 190$~kpc) in the Galaxy, BooIII+Styx will likely prove to be a valuable tracer of the shape of the Galactic potential.

Having identified proper motion outliers from our RV sample that cannot be members of BooIII, we can re-derive the velocity from the remaining candidates (i.e., those listed as members in Table~\ref{tab:members}). We use a maximum-likelihood method \citep[e.g.; ][]{PryorMeylan1993,Hargreaves1994,Kleyna2002} to find $V_{\odot} = 197.1\pm3.6$~km~s$^{-1}$ and $\sigma_{\rm v} = 10.7\pm3.5$~km~s$^{-1}$. This mean velocity is virtually unchanged from that of \citet{Carlin2009}, and the dispersion is consistent with their value of $14.0\pm3.2$~km~s$^{-1}$. Although the dispersion decreased and is now closer to values typical of dwarf galaxies, it is still high, reflecting the apparent disrupting nature of BooIII. We also estimated the mean metallicity and its dispersion, and find [Fe/H]$ = -2.1\pm0.2$ and $\sigma_{\rm [Fe/H]} = 0.55\pm0.19$, virtually unchanged from the previously measured values. However, these are determined from only 10 low $S/N$ spectra of faint stars, so cannot be used to make a definitive statement on the BooIII metallicity. The updated properties of BooIII based on our cleaned sample of members are summarized in Table~\ref{tab:meas_params}.

Because the main sequence turnoff of BooIII is near the magnitude limits of SDSS and PS1, and it has not been followed up with deeper imaging, its structural parameters are poorly constrained. From the known stars, it is possible that BooIII is simply an overdensity of tidal debris among the larger Styx stream. If the luminosity ($M_{\rm V} \approx -5.8$) estimated by \citet{Correnti2009} is correct, then it is reasonable to assume that BooIII is indeed a disrupting dwarf that is the source of the Styx stream, which is broad (like dwarf galaxy streams) and has low surface brightness (i.e., does not likely contain a large fraction of the progenitor's stars). However, if BooIII is instead an overdense portion of the Styx stream (similar to the claims by \citealt{Conn2018CetusII,Conn2018} for Cetus~II and Tucana~V being clumps within the Sagittarius stream and SMC, respectively), our conclusions will not change substantially. The orbit we have measured is derived directly from a co-moving set of stars that are coincident in full 6D phase space. Likewise, our conclusions (based on the velocity and metallicity dispersions) that BooIII derives from a disrupted dwarf would still hold, except that instead of being the progenitor of Styx, it would simply be part of the debris from an unknown progenitor. Of course, if this is the case, then the Styx progenitor must still be unidentified, and our measured orbit can guide future searches for it.

It has been suggested (e.g., \citealt{Pawlowski2012}) that the majority of the Galactic satellites are part of a planar ``Vast Polar Structure'' (VPOS) with thickness $\sim20-30$~kpc, which could arise due to group infall or formation of tidal dwarfs in accretion events. The orbital pole of BooIII's orbit is $(l, b) = (100.9^\circ, -58.9^\circ)$, which is $\sim76^\circ$ from the pole of the VPOS \citep{Pawlowski2013}. Furthermore, our measured proper motions are inconsistent with the predictions by \citet[][their Table~4]{Pawlowski2013} for BooIII's proper motion if it was part of the VPOS. Thus it is unlikely that BooIII is associated with the plane of satellites around the Milky Way.

\begin{figure}
\includegraphics[width=0.95\columnwidth, trim=0.0in 0.0in 0.8in 0.5in]{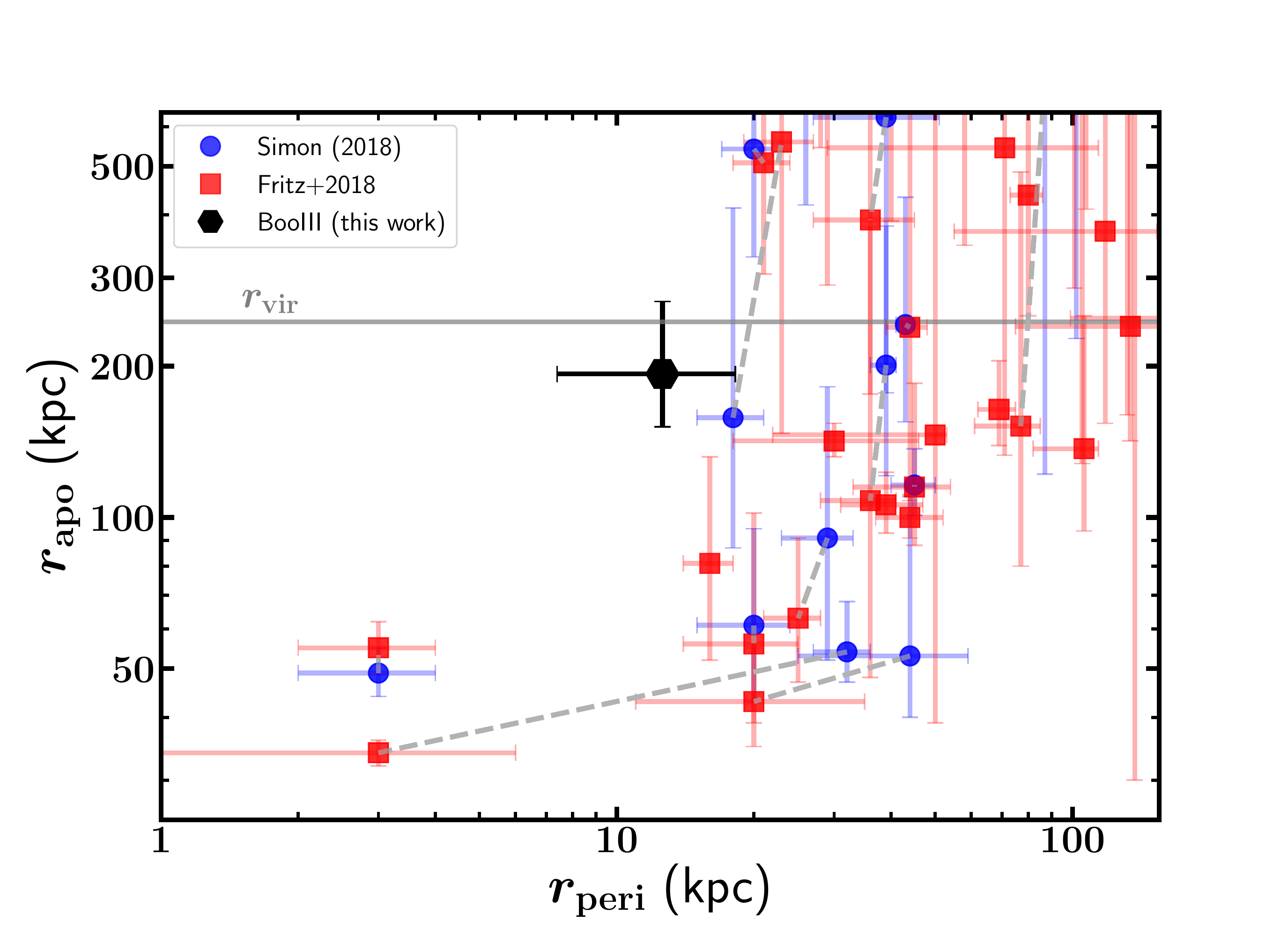}
\caption{Orbital pericenters and apocenters of MW dwarf galaxies as measured by \citet[][blue circles]{Simon2018} and \citet[][red squares]{Fritz18}. Objects in common between the two studies are connected by dashed gray line segments. Our derived parameters for BooIII are shown by the large black hexagon. All measurements reported in this figure used orbits integrated in the \texttt{MWPotential2014} implemented in \texttt{galpy} \citep{Bovy2015galpy}. BooIII occupies a region of parameter space that is not populated among the other MW dwarfs, with a close pericentric passage combined with a large apocenter.}\label{fig:MWdwarfs}
\end{figure}

In Figure~\ref{fig:MWdwarfs}, we compare our measured orbital properties for BooIII to those of dwarf galaxies near the MW (from \citealt{Simon2018, Fritz18}). Only two MW dwarfs have smaller pericenters than BooIII: Tucana~III, which is embedded in an extended stellar stream \citep{DrlicaWagner2015,Li2018}, and the possibly tidally disrupted satellite Segue~2 \citep{Belokurov2009, Kirby2013}, which shows a large discrepancy between the proper motion studies (indeed, the point in Fig.~\ref{fig:MWdwarfs} from \citealt{Fritz18} is derived from only two spectroscopic candidates, and thus fairly unreliable). Both of these objects, while on rather radial orbits, are confined to within $\sim50$~kpc of the Galactic center, while BooIII's orbit extends out to $\sim200$~kpc. The only objects near BooIII in $r_{\rm apo}$-$r_{\rm peri}$ space are Tri~II and Crater~2, which are also possibly remnants of disrupting dwarf galaxies (e.g., \citealt{Kirby2015,Martin2016,Carlin17,Kirby2017} for Tri~II; \citealt{Torrealba16,Sanders2018} -- though \citealt{Caldwell2017} claims no evidence for tidal disruption -- for Crater~2). Based on the lack of satellites with similar orbital parameters, it appears that BooIII may be an object whose orbit would lead to rapid disruption, and that we have happened to catch before it has been completely destroyed. 

We have derived the orbit of the enigmatic stellar overdensity BooIII, and shown it to be consistent with a tidally disrupting dwarf galaxy that is associated with the Styx stream. It is unlikely that BooIII would have survived many pericentric passages on its current radial orbit, so it may have been disrupted on its first infall into the Milky Way halo. BooIII is one of only a few known systems on orbits that pass within $\sim10$~kpc of the Galactic center, and thus provides a valuable probe of the process of tidal disruption and the Galactic potential responsible for that disruption. Deeper imaging and spectroscopic follow-up are needed to fully characterize the nature of BooIII and the associated Styx stream.

\acknowledgments

DJS acknowledges B. Mutlu-Pakdil for providing photometry advice.  Research by DJS is supported by NSF grants AST-1821987 and 1821967. We thank the referee for valuable comments that helped us improve the manuscript.

This work has made use of data from the European Space Agency (ESA) mission
{\it Gaia} (\url{https://www.cosmos.esa.int/gaia}), processed by the {\it Gaia}
Data Processing and Analysis Consortium (DPAC,
\url{https://www.cosmos.esa.int/web/gaia/dpac/consortium}). Funding for the DPAC
has been provided by national institutions, in particular the institutions
participating in the {\it Gaia} Multilateral Agreement.

This research has made use of NASA's Astrophysics Data System, and \texttt{Astropy}, a community-developed core Python package for Astronomy \citep{Astropy}. 

The Pan-STARRS1 Surveys have been made possible through contributions of the Institute for Astronomy, the University of Hawaii, the Pan-STARRS Project Office, the Max-Planck Society and its participating institutes, the Max Planck Institute for Astronomy, Heidelberg and the Max Planck Institute for Extraterrestrial Physics, Garching, The Johns Hopkins University, Durham University, the University of Edinburgh, Queen's University Belfast, the Harvard-Smithsonian Center for Astrophysics, the Las Cumbres Observatory Global Telescope Network Incorporated, the National Central University of Taiwan, the Space Telescope Science Institute, the National Aeronautics and Space Administration under Grant No. NNX08AR22G issued through the Planetary Science Division of the NASA Science Mission Directorate, the National Science Foundation under Grant AST-1238877, the University of Maryland, Eotvos Lorand University (ELTE), and the Los Alamos National Laboratory.

\vspace{5mm}
\facilities{{\it Gaia} DR2, MMT (Hectospec), PS1}

\software{astropy \citep{2013A&A...558A..33A,Astropy}, galpy \citep{Bovy2015galpy}, Matplotlib \citep{Matplotlib}, NumPy \citep{vanderWalt11_numpy}, Topcat \citep{Topcat}.
          }

\bibliographystyle{aasjournal}

\begin{thebibliography}{}
\expandafter\ifx\csname natexlab\endcsname\relax\def\natexlab#1{#1}\fi
\providecommand{\url}[1]{\href{#1}{#1}}


\bibitem[{{Astropy Collaboration} {et~al.}(2013){Astropy Collaboration},
  {Robitaille}, {Tollerud}, {Greenfield}, {Droettboom}, {Bray}, {Aldcroft},
  {Davis}, {Ginsburg}, {Price-Whelan}, {Kerzendorf}, {Conley}, {Crighton},
  {Barbary}, {Muna}, {Ferguson}, {Grollier}, {Parikh}, {Nair}, {Unther},
  {Deil}, {Woillez}, {Conseil}, {Kramer}, {Turner}, {Singer}, {Fox}, {Weaver},
  {Zabalza}, {Edwards}, {Azalee Bostroem}, {Burke}, {Casey}, {Crawford},
  {Dencheva}, {Ely}, {Jenness}, {Labrie}, {Lim}, {Pierfederici}, {Pontzen},
  {Ptak}, {Refsdal}, {Servillat}, \& {Streicher}}]{2013A&A...558A..33A}
{Astropy Collaboration}, {Robitaille}, T.~P., {Tollerud}, E.~J., {et~al.} 2013,
  \aap, 558, A33

\bibitem[{{Belokurov} {et~al.}(2006){Belokurov}, {Evans}, {Irwin}, {Hewett}, \&
  {Wilkinson}}]{Belokurov06}
{Belokurov}, V., {Evans}, N.~W., {Irwin}, M.~J., {Hewett}, P.~C., \&
  {Wilkinson}, M.~I. 2006, \apjl, 637, L29

\bibitem[Belokurov et al.(2009)]{Belokurov2009} Belokurov, V., Walker, M.~G., Evans, N.~W., et al.\ 2009, \mnras, 397, 1748 

\bibitem[{Bernard {et~al.}(2014)Bernard, Ferguson, Schlafly, Platais, Bell,
  Martin, Rix, Slater, Burgett, Chambers, Draper, Hodapp, Kaiser, Kudritzki,
  Magnier, Metcalfe, Tonry, Wainscoat, \& Waters}]{Bernard2014}
Bernard, E.~J., Ferguson, A. M.~N., Schlafly, E.~F., {et~al.} 2014, Mon. Not.
  R. Astron. Soc., 2999

\bibitem[{{Besla} {et~al.}(2007){Besla}, {Kallivayalil}, {Hernquist},
  {Robertson}, {Cox}, {van der Marel}, \& {Alcock}}]{Besla07}
{Besla}, G., {Kallivayalil}, N., {Hernquist}, L., {et~al.} 2007, \apj, 668, 949

\bibitem[{{Bland-Hawthorn} \& {Gerhard}(2016)}]{BlandHawthornGerhard2016}
{Bland-Hawthorn}, J., \& {Gerhard}, O. 2016, \araa, 54, 529

\bibitem[{Bovy(2015)}]{Bovy2015galpy}
Bovy, J. 2015, The Astrophysical Journal Supplement Series, 216, 29

\bibitem[{Brown {et~al.}(2018)Brown, Vallenari, Prusti, \&
  de~Bruijne}]{GaiaDR2}
Brown, A. G.~A., Vallenari, A., Prusti, T., \& de~Bruijne, J. H.~J. 2018,
  Astronomy {\&} Astrophysics, doi:10.1051/0004-6361/201833051

\bibitem[Caldwell et al.(2017)]{Caldwell2017} Caldwell, N., Walker, M.~G., Mateo, M., et al.\ 2017, \apj, 839, 20 

\bibitem[{Carlin {et~al.}(2009)Carlin, Grillmair, Mu{\~{n}}oz, Nidever, \&
  Majewski}]{Carlin2009}
Carlin, J.~L., Grillmair, C.~J., Mu{\~{n}}oz, R.~R., Nidever, D.~L., \&
  Majewski, S.~R. 2009, The Astrophysical Journal, 702, L9

\bibitem[{{Carlin} {et~al.}(2017){Carlin}, {Sand}, {Mu{\~n}oz}, {Spekkens},
  {Willman}, {Crnojevi{\'c}}, {Forbes}, {Hargis}, {Kirby}, {Peter},
  {Romanowsky}, \& {Strader}}]{Carlin17}
{Carlin}, J.~L., {Sand}, D.~J., {Mu{\~n}oz}, R.~R., {et~al.} 2017, \aj, 154,
  267

\bibitem[{{Carretta} {et~al.}(2009){Carretta}, {Bragaglia}, {Gratton},
  {Lucatello}, {Catanzaro}, {Leone}, {Bellazzini}, {Claudi}, {D'Orazi},
  {Momany}, {Ortolani}, {Pancino}, {Piotto}, {Recio-Blanco}, \&
  {Sabbi}}]{Carretta09}
{Carretta}, E., {Bragaglia}, A., {Gratton}, R.~G., {et~al.} 2009, \aap, 505,
  117

\bibitem[{{Casetti-Dinescu} \& {Girard}(2016)}]{Dana16}
{Casetti-Dinescu}, D.~I., \& {Girard}, T.~M. 2016, \mnras, 461, 271

\bibitem[{{Casetti-Dinescu} {et~al.}(2018){Casetti-Dinescu}, {Girard}, \&
  {Schriefer}}]{Dana18}
{Casetti-Dinescu}, D.~I., {Girard}, T.~M., \& {Schriefer}, M. 2018, \mnras,
  473, 4064

\bibitem[{{Chambers} {et~al.}(2016){Chambers}, {Magnier}, {Metcalfe},
  {Flewelling}, {Huber}, {Waters}, {Denneau}, {Draper}, {Farrow}, {Finkbeiner},
  {Holmberg}, {Koppenhoefer}, {Price}, {Saglia}, {Schlafly}, {Smartt},
  {Sweeney}, {Wainscoat}, {Burgett}, {Grav}, {Heasley}, {Hodapp}, {Jedicke},
  {Kaiser}, {Kudritzki}, {Luppino}, {Lupton}, {Monet}, {Morgan}, {Onaka},
  {Stubbs}, {Tonry}, {Banados}, {Bell}, {Bender}, {Bernard}, {Botticella},
  {Casertano}, {Chastel}, {Chen}, {Chen}, {Cole}, {Deacon}, {Frenk},
  {Fitzsimmons}, {Gezari}, {Goessl}, {Goggia}, {Goldman}, {Grebel}, {Hambly},
  {Hasinger}, {Heavens}, {Heckman}, {Henderson}, {Henning}, {Holman}, {Hopp},
  {Ip}, {Isani}, {Keyes}, {Koekemoer}, {Kotak}, {Long}, {Lucey}, {Liu},
  {Martin}, {McLean}, {Morganson}, {Murphy}, {Nieto-Santisteban}, {Norberg},
  {Peacock}, {Pier}, {Postman}, {Primak}, {Rae}, {Rest}, {Riess}, {Riffeser},
  {Rix}, {Roser}, {Schilbach}, {Schultz}, {Scolnic}, {Szalay}, {Seitz},
  {Shiao}, {Small}, {Smith}, {Soderblom}, {Taylor}, {Thakar}, {Thiel},
  {Thilker}, {Urata}, {Valenti}, {Walter}, {Watters}, {Werner}, {White},
  {Wood-Vasey}, \& {Wyse}}]{Chambers2016_PS1}
{Chambers}, K.~C., {Magnier}, E.~A., {Metcalfe}, N., {et~al.} 2016, ArXiv
  e-prints, arXiv:1612.05560

\bibitem[{{Collins} {et~al.}(2017){Collins}, {Tollerud}, {Sand}, {Bonaca},
  {Willman}, \& {Strader}}]{Collins17}
{Collins}, M.~L.~M., {Tollerud}, E.~J., {Sand}, D.~J., {et~al.} 2017, \mnras,
  467, 573

\bibitem[{{Conn} {et~al.}(2018{\natexlab{a}}){Conn}, {Jerjen}, {Kim}, \&
  {Schirmer}}]{Conn2018CetusII}
{Conn}, B.~C., {Jerjen}, H., {Kim}, D., \& {Schirmer}, M. 2018{\natexlab{a}},
  \apj, 857, 70

\bibitem[{{Conn} {et~al.}(2018{\natexlab{b}}){Conn}, {Jerjen}, {Kim}, \&
  {Schirmer}}]{Conn2018}
---. 2018{\natexlab{b}}, \apj, 852, 68

\bibitem[{{Correnti} {et~al.}(2009){Correnti}, {Bellazzini}, \&
  {Ferraro}}]{Correnti2009}
{Correnti}, M., {Bellazzini}, M., \& {Ferraro}, F.~R. 2009, \mnras, 397, L26

\bibitem[{{D'Onghia} \& {Fox}(2016)}]{D16}
{D'Onghia}, E., \& {Fox}, A.~J. 2016, \araa, 54, 363

\bibitem[Drlica-Wagner et al.(2015)]{DrlicaWagner2015} Drlica-Wagner, A., Bechtol, K., Rykoff, E.~S., et al.\ 2015, \apj, 813, 109 

\bibitem[{{Flewelling} {et~al.}(2016){Flewelling}, {Magnier}, {Chambers},
  {Heasley}, {Holmberg}, {Huber}, {Sweeney}, {Waters}, {Chen}, {Farrow},
  {Hasinger}, {Henderson}, {Long}, {Metcalfe}, {Nieto-Santisteban}, {Norberg},
  {Saglia}, {Szalay}, {Rest}, {Thakar}, {Tonry}, {Valenti}, {Werner}, {White},
  {Denneau}, {Draper}, {Hodapp}, {Jedicke}, {Kaiser}, {Kudritzki}, {Price},
  {Wainscoat}, {Chastel}, {McClean}, {Postman}, \& {Shiao}}]{Flewelling2016}
{Flewelling}, H.~A., {Magnier}, E.~A., {Chambers}, K.~C., {et~al.} 2016, ArXiv
  e-prints, arXiv:1612.05243

\bibitem[{{Fritz} {et~al.}(2018){Fritz}, {Battaglia}, {Pawlowski},
  {Kallivayalil}, {van der Marel}, {Sohn}, {Brook}, \& {Besla}}]{Fritz18}
{Fritz}, T.~K., {Battaglia}, G., {Pawlowski}, M.~S., {et~al.} 2018, ArXiv
  e-prints, arXiv:1805.00908

\bibitem[{{Fritz} {et~al.}(2017){Fritz}, {Lokken}, {Kallivayalil}, {Wetzel},
  {Linden}, {Zivick}, \& {Tollerud}}]{Fritz17}
{Fritz}, T.~K., {Lokken}, M., {Kallivayalil}, N., {et~al.} 2017, ArXiv
  e-prints, arXiv:1711.09097

\bibitem[{{Gaia Collaboration} {et~al.}(2018){Gaia Collaboration}, {Helmi},
  {van Leeuwen}, {McMillan}, {Massari}, {Antoja}, {Robin}, {Lindegren},
  {Bastian}, \& {co-authors}}]{Helmi18}
{Gaia Collaboration}, {Helmi}, A., {van Leeuwen}, F., {et~al.} 2018, ArXiv
  e-prints, arXiv:1804.09381

\bibitem[{{Garling} {et~al.}(2018){Garling}, {Willman}, {Sand}, {Hargis},
  {Crnojevi{\'c}}, {Bechtol}, {Carlin}, {Strader}, {Zou}, {Zhou}, {Nie},
  {Zhang}, {Zhou}, \& {Peng}}]{Garling18}
{Garling}, C., {Willman}, B., {Sand}, D.~J., {et~al.} 2018, \apj, 852, 44

\bibitem[{Grillmair(2009)}]{Grillmair2009}
Grillmair, C.~J. 2009, The Astrophysical Journal, 693, 1118

\bibitem[{{Grillmair} \& {Johnson}(2006)}]{Grillmair06}
{Grillmair}, C.~J., \& {Johnson}, R. 2006, \apjl, 639, L17

\bibitem[{{Hargreaves} {et~al.}(1994){Hargreaves}, {Gilmore}, {Irwin}, \&
  {Carter}}]{Hargreaves1994}
{Hargreaves}, J.~C., {Gilmore}, G., {Irwin}, M.~J., \& {Carter}, D. 1994,
  \mnras, 269, 957

\bibitem[{{Harris}(1996)}]{Harriscat}
{Harris}, W.~E. 1996, \aj, 112, 1487

\bibitem[{Helmi {et~al.}(2018)Helmi, van Leeuwen, McMillan, \&
  DPAC}]{Helmi2018}
Helmi, A., van Leeuwen, F., McMillan, P., \& DPAC. 2018, Astronomy {\&}
  Astrophysics, doi:10.1051/0004-6361/201832698

\bibitem[{Hunter(2007)}]{Matplotlib}
Hunter, J.~D. 2007, Computing In Science \& Engineering, 9, 90

\bibitem[{{Kallivayalil} {et~al.}(2006){Kallivayalil}, {van der Marel},
  {Alcock}, {Axelrod}, {Cook}, {Drake}, \& {Geha}}]{Kall06}
{Kallivayalil}, N., {van der Marel}, R.~P., {Alcock}, C., {et~al.} 2006, \apj,
  638, 772

\bibitem[{{Kallivayalil} {et~al.}(2013){Kallivayalil}, {van der Marel},
  {Besla}, {Anderson}, \& {Alcock}}]{Kall13}
{Kallivayalil}, N., {van der Marel}, R.~P., {Besla}, G., {Anderson}, J., \&
  {Alcock}, C. 2013, \apj, 764, 161

\bibitem[{{Kallivayalil} {et~al.}(2018){Kallivayalil}, {Sales}, {Zivick},
  {Fritz}, {Del Pino}, {Sohn}, {Besla}, {van der Marel}, {Navarro}, \&
  {Sacchi}}]{Kall18}
{Kallivayalil}, N., {Sales}, L., {Zivick}, P., {et~al.} 2018, ArXiv e-prints,
  arXiv:1805.01448

\bibitem[Kirby et al.(2013)]{Kirby2013} Kirby, E.~N., Boylan-Kolchin, M., Cohen, J.~G., et al.\ 2013, \apj, 770, 16 

\bibitem[Kirby et al.(2015)]{Kirby2015} Kirby, E.~N., Cohen, J.~G., Simon, J.~D., \& Guhathakurta, P.\ 2015, \apjl, 814, L7 

\bibitem[Kirby et al.(2017)]{Kirby2017} Kirby, E.~N., Cohen, J.~G., Simon, J.~D., et al.\ 2017, \apj, 838, 83 


\bibitem[{{Kleyna} {et~al.}(2002){Kleyna}, {Wilkinson}, {Evans}, {Gilmore}, \&
  {Frayn}}]{Kleyna2002}
{Kleyna}, J., {Wilkinson}, M.~I., {Evans}, N.~W., {Gilmore}, G., \& {Frayn}, C.
  2002, \mnras, 330, 792

\bibitem[{{Law} {et~al.}(2009){Law}, {Kulkarni}, {Dekany}, {Ofek}, {Quimby},
  {Nugent}, {Surace}, {Grillmair}, {Bloom}, {Kasliwal}, {Bildsten}, {Brown},
  {Cenko}, {Ciardi}, {Croner}, {Djorgovski}, {van Eyken}, {Filippenko}, {Fox},
  {Gal-Yam}, {Hale}, {Hamam}, {Helou}, {Henning}, {Howell}, {Jacobsen},
  {Laher}, {Mattingly}, {McKenna}, {Pickles}, {Poznanski}, {Rahmer}, {Rau},
  {Rosing}, {Shara}, {Smith}, {Starr}, {Sullivan}, {Velur}, {Walters}, \&
  {Zolkower}}]{Law2009PS1}
{Law}, N.~M., {Kulkarni}, S.~R., {Dekany}, R.~G., {et~al.} 2009, \pasp, 121,
  1395

\bibitem[Li et al.(2018)]{Li2018} Li, T.~S., Simon, J.~D., Kuehn, K., et al.\ 2018, arXiv:1804.07761 

\bibitem[{{Magnier} {et~al.}(2016){Magnier}, {Schlafly}, {Finkbeiner}, {Tonry},
  {Goldman}, {R{\"o}ser}, {Schilbach}, {Chambers}, {Flewelling}, {Huber},
  {Price}, {Sweeney}, {Waters}, {Denneau}, {Draper}, {Hodapp}, {Jedicke},
  {Kudritzki}, {Metcalfe}, {Stubbs}, \& {Wainscoast}}]{Magnier2016}
{Magnier}, E.~A., {Schlafly}, E.~F., {Finkbeiner}, D.~P., {et~al.} 2016, ArXiv
  e-prints, arXiv:1612.05242

\bibitem[Martin et al.(2016)]{Martin2016} Martin, N.~F., Ibata, R.~A., Collins, M.~L.~M., et al.\ 2016, \apj, 818, 40 

\bibitem[{{Massari} \& {Helmi}(2018)}]{Massari18}
{Massari}, D., \& {Helmi}, A. 2018, ArXiv e-prints, arXiv:1805.01839

\bibitem[{{Mu{\~n}oz} {et~al.}(2010){Mu{\~n}oz}, {Geha}, \&
  {Willman}}]{Munoz10}
{Mu{\~n}oz}, R.~R., {Geha}, M., \& {Willman}, B. 2010, \aj, 140, 138

\bibitem[{{Mutlu-Pakdil} {et~al.}(2018){Mutlu-Pakdil}, {Sand}, {Carlin},
  {Spekkens}, {Caldwell}, {Crnojevi{\'c}}, {Hughes}, {Willman}, \&
  {Zaritsky}}]{Mutlu18}
{Mutlu-Pakdil}, B., {Sand}, D.~J., {Carlin}, J.~L., {et~al.} 2018, ArXiv
  e-prints, arXiv:1804.08627

\bibitem[{{Navarro} {et~al.}(1997){Navarro}, {Frenk}, \& {White}}]{NFW1997}
{Navarro}, J.~F., {Frenk}, C.~S., \& {White}, S. D.~M. 1997, \apj, 490, 493

\bibitem[{Pawlowski \& Kroupa(2013)}]{Pawlowski2013}
Pawlowski, M.~S., \& Kroupa, P. 2013, Monthly Notices of the Royal Astronomical
  Society, 435, 2116

\bibitem[{{Pawlowski} {et~al.}(2012){Pawlowski}, {Pflamm-Altenburg}, \&
  {Kroupa}}]{Pawlowski2012}
{Pawlowski}, M.~S., {Pflamm-Altenburg}, J., \& {Kroupa}, P. 2012, \mnras, 423,
  1109

\bibitem[{{Piatek} {et~al.}(2016){Piatek}, {Pryor}, \& {Olszewski}}]{Piatek16}
{Piatek}, S., {Pryor}, C., \& {Olszewski}, E.~W. 2016, \aj, 152, 166

\bibitem[{{Pryor} \& {Meylan}(1993)}]{PryorMeylan1993}
{Pryor}, C., \& {Meylan}, G. 1993, in Astronomical Society of the Pacific
  Conference Series, Vol.~50, Structure and Dynamics of Globular Clusters, ed.
  S.~G. {Djorgovski} \& G.~{Meylan}, 357

\bibitem[{{Pryor} {et~al.}(2015){Pryor}, {Piatek}, \& {Olszewski}}]{Pryor15}
{Pryor}, C., {Piatek}, S., \& {Olszewski}, E.~W. 2015, \aj, 149, 42

\bibitem[{{Rau} {et~al.}(2009){Rau}, {Kulkarni}, {Law}, {Bloom}, {Ciardi},
  {Djorgovski}, {Fox}, {Gal-Yam}, {Grillmair}, {Kasliwal}, {Nugent}, {Ofek},
  {Quimby}, {Reach}, {Shara}, {Bildsten}, {Cenko}, {Drake}, {Filippenko},
  {Helfand}, {Helou}, {Howell}, {Poznanski}, \& {Sullivan}}]{Rau2009PS1}
{Rau}, A., {Kulkarni}, S.~R., {Law}, N.~M., {et~al.} 2009, \pasp, 121, 1334

\bibitem[{{Roderick} {et~al.}(2015){Roderick}, {Jerjen}, {Mackey}, \& {Da
  Costa}}]{Roderick15}
{Roderick}, T.~A., {Jerjen}, H., {Mackey}, A.~D., \& {Da Costa}, G.~S. 2015,
  \apj, 804, 134

\bibitem[{{Sand} {et~al.}(2009){Sand}, {Olszewski}, {Willman}, {Zaritsky},
  {Seth}, {Harris}, {Piatek}, \& {Saha}}]{Sand09}
{Sand}, D.~J., {Olszewski}, E.~W., {Willman}, B., {et~al.} 2009, \apj, 704, 898

\bibitem[{{Sand} {et~al.}(2012){Sand}, {Strader}, {Willman}, {Zaritsky},
  {McLeod}, {Caldwell}, {Seth}, \& {Olszewski}}]{Sand12}
{Sand}, D.~J., {Strader}, J., {Willman}, B., {et~al.} 2012, \apj, 756, 79

\bibitem[Sanders et al.(2018)]{Sanders2018} Sanders, J.~L., Evans, N.~W., \& Dehnen, W.\ 2018, \mnras, 478, 3879 

\bibitem[{{Schlafly} \& {Finkbeiner}(2011)}]{Schlafly2011}
{Schlafly}, E.~F., \& {Finkbeiner}, D.~P. 2011, \apj, 737, 103

\bibitem[{{Schlegel} {et~al.}(1998){Schlegel}, {Finkbeiner}, \&
  {Davis}}]{Schlegel1998}
{Schlegel}, D.~J., {Finkbeiner}, D.~P., \& {Davis}, M. 1998, \apj, 500, 525

\bibitem[{{Sch{\"o}nrich} {et~al.}(2010){Sch{\"o}nrich}, {Binney}, \&
  {Dehnen}}]{Schoenrich2010}
{Sch{\"o}nrich}, R., {Binney}, J., \& {Dehnen}, W. 2010, \mnras, 403, 1829

\bibitem[{Sesar {et~al.}(2014)Sesar, Banholzer, Cohen, Martin, Grillmair,
  Levitan, Laher, Ofek, Surace, Kulkarni, Prince, \& Rix}]{Sesar2014}
Sesar, B., Banholzer, S.~R., Cohen, J.~G., {et~al.} 2014, Astrophys. J., 793,
  135

\bibitem[{Simon(2018)}]{Simon2018}
Simon, J.~D. 2018, arXiv:1804.10230

\bibitem[{{Sohn} {et~al.}(2013){Sohn}, {Besla}, {van der Marel},
  {Boylan-Kolchin}, {Majewski}, \& {Bullock}}]{Sohn13}
{Sohn}, S.~T., {Besla}, G., {van der Marel}, R.~P., {et~al.} 2013, \apj, 768,
  139

\bibitem[{{Sohn} {et~al.}(2017){Sohn}, {Patel}, {Besla}, {van der Marel},
  {Bullock}, {Strigari}, {van de Ven}, {Walker}, \& {Bellini}}]{Sohn17}
{Sohn}, S.~T., {Patel}, E., {Besla}, G., {et~al.} 2017, \apj, 849, 93

\bibitem[{{Taylor}(2005)}]{Topcat}
{Taylor}, M.~B. 2005, in Astronomical Society of the Pacific Conference Series,
  Vol. 347, Astronomical Data Analysis Software and Systems XIV, ed.
  P.~{Shopbell}, M.~{Britton}, \& R.~{Ebert}, 29

\bibitem[{{The Astropy Collaboration} {et~al.}(2018){The Astropy
  Collaboration}, {Price-Whelan}, {Sip{\H o}cz}, {G{\"u}nther}, {Lim},
  {Crawford}, {Conseil}, {Shupe}, {Craig}, {Dencheva}, {Ginsburg},
  {VanderPlas}, {Bradley}, {P{\'e}rez-Su{\'a}rez}, {de Val-Borro}, {Aldcroft},
  {Cruz}, {Robitaille}, {Tollerud}, {Ardelean}, {Babej}, {Bachetti}, {Bakanov},
  {Bamford}, {Barentsen}, {Barmby}, {Baumbach}, {Berry}, {Biscani}, {Boquien},
  {Bostroem}, {Bouma}, {Brammer}, {Bray}, {Breytenbach}, {Buddelmeijer},
  {Burke}, {Calderone}, {Cano Rodr{\'{\i}}guez}, {Cara}, {Cardoso},
  {Cheedella}, {Copin}, {Crichton}, {D{\'A}vella}, {Deil}, {Depagne},
  {Dietrich}, {Donath}, {Droettboom}, {Earl}, {Erben}, {Fabbro}, {Ferreira},
  {Finethy}, {Fox}, {Garrison}, {Gibbons}, {Goldstein}, {Gommers}, {Greco},
  {Greenfield}, {Groener}, {Grollier}, {Hagen}, {Hirst}, {Homeier}, {Horton},
  {Hosseinzadeh}, {Hu}, {Hunkeler}, {Ivezi{\'c}}, {Jain}, {Jenness}, {Kanarek},
  {Kendrew}, {Kern}, {Kerzendorf}, {Khvalko}, {King}, {Kirkby}, {Kulkarni},
  {Kumar}, {Lee}, {Lenz}, {Littlefair}, {Ma}, {Macleod}, {Mastropietro},
  {McCully}, {Montagnac}, {Morris}, {Mueller}, {Mumford}, {Muna}, {Murphy},
  {Nelson}, {Nguyen}, {Ninan}, {N{\"o}the}, {Ogaz}, {Oh}, {Parejko}, {Parley},
  {Pascual}, {Patil}, {Patil}, {Plunkett}, {Prochaska}, {Rastogi}, {Reddy
  Janga}, {Sabater}, {Sakurikar}, {Seifert}, {Sherbert}, {Sherwood-Taylor},
  {Shih}, {Sick}, {Silbiger}, {Singanamalla}, {Singer}, {Sladen}, {Sooley},
  {Sornarajah}, {Streicher}, {Teuben}, {Thomas}, {Tremblay}, {Turner},
  {Terr{\'o}n}, {van Kerkwijk}, {de la Vega}, {Watkins}, {Weaver}, {Whitmore},
  {Woillez}, \& {Zabalza}}]{Astropy}
{The Astropy Collaboration}, {Price-Whelan}, A.~M., {Sip{\H o}cz}, B.~M.,
  {et~al.} 2018, ArXiv e-prints, arXiv:1801.02634

\bibitem[{{Torrealba} {et~al.}(2016){Torrealba}, {Koposov}, {Belokurov}, \&
  {Irwin}}]{Torrealba16}
{Torrealba}, G., {Koposov}, S.~E., {Belokurov}, V., \& {Irwin}, M. 2016,
  \mnras, 459, 2370

\bibitem[{van~der Walt {et~al.}(2011)van~der Walt, Colbert, \&
  Varoquaux}]{vanderWalt11_numpy}
van~der Walt, S., Colbert, S.~C., \& Varoquaux, G. 2011, Computing in Science
  Engineering, 13, 22

\bibitem[{{Watkins} {et~al.}(2018){Watkins}, {van der Marel}, {Sohn}, \&
  {Evans}}]{Watkins18}
{Watkins}, L.~L., {van der Marel}, R.~P., {Sohn}, S.~T., \& {Evans}, N.~W.
  2018, ArXiv e-prints, arXiv:1804.11348

\end{thebibliography}

\clearpage
\begin{table*}
\caption{Properties of Bo\"otes III RV sample.}\label{tab:members}
\tiny
\begin{tabular}{ccccccccccccc}
\hline
\hline
id & source$\_$id & $\alpha$ & $\delta$ & $g$ & $r$ & $i$ & $v_{\rm helio}$ & $\mu_{\alpha*}$ & $\mu_\delta$ & [Fe/H]\tablenotemark{a} & D\tablenotemark{b} & member? \\
 & ({\it Gaia} DR2) & deg & deg & mag & mag & mag & $\mathrm{km\,s^{-1}}$ & $\mathrm{mas\,yr^{-1}}$ & $\mathrm{mas\,yr^{-1}}$ &  & arcmin & \\
\hline
s135549+263008 & 1450811159128205568 & 208.955653 & 26.502208 & 21.01 & 20.67 & 20.45 & 195.5$\pm$11.4 & 1.078$\pm$3.305 & 5.259$\pm$3.285 & -3.3 & 23.9 & N \\
s135603+263942 & 1450817000283805824 & 209.013621 & 26.661679 & 20.40 & 19.95 & 19.78 & 204.0$\pm$8.5 & -0.466$\pm$1.308 & -0.583$\pm$1.047 & -2.5 & 15.9 & Y\\
s135630+265024 & 1450826552290908800 & 209.128453 & 26.840015 & 18.96 & 19.05 & 19.17 & 171.6$\pm$11.1 & -1.211$\pm$0.562 & -1.139$\pm$0.436 & --- & 9.0 & Y \\
s135638+265222 & 1450826724089602816 & 209.159600 & 26.872888 & 20.19 & 19.73 & 19.56 & 188.9$\pm$7.7 & -1.84$\pm$0.844 & -1.784$\pm$0.761 & -2.3 & 8.8 & Y \\
s135642+265847 & 1451038792395234560 & 209.174766 & 26.979976 & 18.94 & 19.05 & 19.17 & 180.4$\pm$10.9 & -1.932$\pm$0.508 & 0.579$\pm$0.5 & --- & 13.5 & Y \\
s135702+264947 & 1450823051893369984 & 209.261163 & 26.829991 & 20.03 & 19.61 & 19.40 & 197.8$\pm$7.2 & -1.709$\pm$0.767 & -0.332$\pm$0.631 & -2.7 & 3.5 & Y\\
s135719+263424 & 1450803222028673280 & 209.331958 & 26.573608 & 19.10 & 19.30 & 19.50 & 220.8$\pm$12.3 & -1.492$\pm$0.531 & -0.928$\pm$0.48 & --- & 12.4 & Y\\
s135752+264815 & 1450833767836315392 & 209.467272 & 26.804392 & 18.99 & 19.16 & 19.35 & 209.3$\pm$11.5 & -0.207$\pm$0.708 & -0.927$\pm$0.735 & --- & 10.1 & Y\\
s135755+263953 & 1450828682594980736 & 209.479318 & 26.664838 & 20.16 & 19.71 & 19.52 & 185.7$\pm$7.7 & -0.462$\pm$0.786 & -0.479$\pm$0.673 & -2.1 & 12.5 & Y\\
s135800+265009 & 1450835314024556544 & 209.500251 & 26.835870 & 20.99 & 20.61 & 20.47 & 195.5$\pm$11.0 & -1.006$\pm$2.148 & -1.389$\pm$1.696 & -1.0 & 12.3 & Y\\
s135804+261716 & 1450748486965605760 & 209.520145 & 26.287895 & 18.94 & 19.07 & 19.24 & 185.0$\pm$10.9 & -0.566$\pm$0.515 & -1.793$\pm$0.441 & --- & 31.9 & Y\\
s135817+265641 & 1450842185972291840 & 209.573497 & 26.944967 & 20.52 & 20.16 & 19.94 & 210.4$\pm$9.2 & -1.435$\pm$0.994 & 0.541$\pm$0.904 & -3.2 & 18.7 & Y\\
s135826+263904 & 1450781163077088000 & 209.610213 & 26.651188 & 20.30 & 19.92 & 19.71 & 199.6$\pm$7.9 & 0.058$\pm$0.988 & -1.044$\pm$0.84 & -2.2 & 19.2 & Y\\
s135840+264242 & 1450782949783229824 & 209.668856 & 26.711657 & 20.31 & 20.11 & 19.99 & 191.2$\pm$8.2 & -2.084$\pm$1.058 & -2.503$\pm$0.853 & -1.4 & 21.1 & Y\\
s135843+262802 & 1450751617997283840 & 209.681364 & 26.467320 & 20.19 & 19.90 & 19.77 & 166.4$\pm$7.5 & -5.797$\pm$0.953 & -2.153$\pm$0.944 & -2.1 & 28.3 & N\\
s135848+262425 & 1450739381635432192 & 209.700291 & 26.407019 & 20.20 & 19.89 & 19.77 & 223.8$\pm$7.7 & 0.24$\pm$0.906 & -1.969$\pm$0.914 & -0.9 & 31.5 & Y\\
s135914+264136 & 1450785045727265792 & 209.810712 & 26.693513 & 18.93 & 19.04 & 19.15 & 209.2$\pm$10.8 & -1.595$\pm$0.487 & -1.315$\pm$0.436 & --- & 28.8 & Y\\
s135922+264511 & 1450788584780752256 & 209.845448 & 26.753301 & 20.49 & 20.13 & 19.99 & 221.2$\pm$9.0 & -7.457$\pm$1.351 & -2.497$\pm$1.315 & -1.4 & 30.3 & N\\
s135940+263918 & 1450783774417361280 & 209.916855 & 26.655120 & 20.86 & 20.39 & 20.24 & 176.3$\pm$10.2 & -10.529$\pm$1.368 & -11.886$\pm$1.454 & -1.9 & 34.8 & N\\
s135757+265130 & -9223372036854775808 & 209.489916 & 26.858467 & 21.88 & 21.46 & 21.14 & 206.6$\pm$16.0 & N/A\tablenotemark{d} & N/A & -0.9 & 12.3 & N \\
BooIII$\_$RR1\tablenotemark{c} & 1258556500130302080 & 210.143865 & 25.931296 & 19.22 & 19.01 & 18.96 & 173.0$\pm$13.0 & -1.34$\pm$0.417 & -0.978$\pm$0.377 & -2.0 & 68.7 & Y \\
\hline
\end{tabular}
\tablenotetext{}{Positions and proper motions from {\it Gaia} DR2, and $gri$ magnitudes are from PanSTARRS-1. IDs, radial velocities and [Fe/H] are from \citet{Carlin2009}. }
\tablenotetext{a}{For RGB and RR Lyrae stars only; metallicities were not reported by \citet{Carlin2009} for BHB stars.}
\tablenotetext{b}{Angular distance from the BooIII center derived by \citet{Grillmair2009}.}
\tablenotetext{c}{RR Lyrae star from \citet{Sesar2014}.}
\tablenotetext{d}{No match in the {\it Gaia} database.}

\end{table*}

\clearpage
\begin{table}
\begin{center}
\caption{Properties of proper-motion and isochrone-selected candidate members of Bo\"otes III.}\label{tab:pm_cands}
\footnotesize
\begin{tabular}{ccccccccc}
\hline
\hline
source$\_$id & $\alpha$ & $\delta$ & $g$ & $r$ & $i$ & $\mu_{\alpha*}$ & $\mu_\delta$ & D\tablenotemark{a} \\
({\it Gaia} DR2) & deg & deg & mag & mag & mag & $\mathrm{mas\,yr^{-1}}$ & $\mathrm{mas\,yr^{-1}}$ & arcmin\\
\hline
1450755118394910720 & 209.529019 & 26.471237 & 18.85 & 18.78 & 18.82 & -1.23$\pm$0.462 & -0.131$\pm$0.414 & 22.6 \\
1450750170592551040 & 209.529158 & 26.385552 & 18.86 & 18.75 & 18.80 & -2.088$\pm$0.387 & -0.468$\pm$0.364 & 26.9 \\
1450794460295316224 & 209.172508 & 26.415167 & 19.97 & 19.67 & 19.27 & -0.02$\pm$0.589 & -0.587$\pm$0.661 & 22.4 \\
1450793433798851328 & 209.204334 & 26.378583 & 19.52 & 18.91 & 18.70 & -1.719$\pm$0.428 & -1.247$\pm$0.426 & 24.1 \\
1451010170732704896 & 208.823004 & 26.797716 & 18.57 & 18.05 & 17.84 & -1.261$\pm$0.248 & -1.461$\pm$0.242 & 24.6 \\
1451011923079404544 & 208.854035 & 26.905210 & 18.85 & 18.90 & 18.98 & -1.032$\pm$0.406 & -0.8$\pm$0.397 & 24.1 \\
1450435091791688448 & 208.890655 & 26.424760 & 19.12 & 18.52 & 18.31 & 0.138$\pm$0.324 & -2.964$\pm$0.301 & 29.7 \\
1451003917260608384 & 208.836855 & 26.663041 & 19.90 & 19.55 & 19.39 & -2.176$\pm$0.885 & -0.996$\pm$0.762 & 24.7 \\
1450820608056155648 & 209.204661 & 26.736842 & 19.74 & 19.29 & 19.08 & -0.917$\pm$0.578 & -0.51$\pm$0.472 & 4.7 \\
1450752026018903168 & 209.556687 & 26.444826 & 18.80 & 18.31 & 18.08 & -1.435$\pm$0.287 & -0.534$\pm$0.281 & 24.7 \\
1450830267438457344 & 209.612310 & 26.759489 & 19.42 & 18.95 & 18.74 & -0.922$\pm$0.476 & -0.931$\pm$0.381 & 17.8 \\
1450782713560551808 & 209.667235 & 26.695234 & 16.31 & 15.36 & 14.92 & -1.199$\pm$0.065 & -0.786$\pm$0.058 & 21.3 \\
1450809617235662336 & 209.382053 & 26.737508 & 17.83 & 17.12 & 16.81 & -0.855$\pm$0.277 & -1.699$\pm$0.252 & 5.9 \\
1450756767662816768 & 209.526091 & 26.552868 & 18.39 & 17.92 & 17.69 & -1.292$\pm$0.224 & -0.909$\pm$0.203 & 18.7 \\
1450757042540728960 & 209.504614 & 26.577302 & 19.38 & 18.93 & 18.67 & -1.269$\pm$0.445 & -0.69$\pm$0.396 & 16.9 \\
1451043602758623232 & 209.092316 & 27.092680 & 18.49 & 17.84 & 17.55 & -0.695$\pm$0.313 & -1.039$\pm$0.28 & 21.6 \\
1451046076659819776 & 209.245742 & 27.212404 & 18.91 & 19.00 & 19.16 & -2.493$\pm$0.521 & -1.649$\pm$0.411 & 26.3 \\
1451047760286978176 & 209.342306 & 27.142651 & 18.04 & 17.39 & 17.13 & 0.353$\pm$0.189 & -1.49$\pm$0.155 & 22.3 \\
1451047794646719104 & 209.354036 & 27.153785 & 19.36 & 18.91 & 18.74 & -2.63$\pm$0.476 & -2.82$\pm$0.405 & 23.1 \\
1451039548309484288 & 209.215315 & 27.031907 & 19.11 & 19.25 & 18.89 & 0.571$\pm$0.542 & -0.179$\pm$0.479 & 15.8 \\
1450858678646760192 & 209.435281 & 27.067580 & 18.92 & 18.72 & 18.66 & -1.641$\pm$0.392 & -1.876$\pm$0.322 & 19.4 \\
1450833596037626752 & 209.396603 & 26.812921 & 18.44 & 17.90 & 17.63 & -1.152$\pm$0.23 & -0.716$\pm$0.215 & 6.6 \\
1450843049261137408 & 209.689069 & 26.969028 & 18.79 & 18.07 & 18.10 & -0.356$\pm$0.245 & 0.497$\pm$0.232 & 24.8 \\
\hline
\end{tabular}
\tablenotetext{}{Positions and proper motions from {\it Gaia} DR2, and $gri$ magnitudes are from PanSTARRS-1.}
\tablenotetext{a}{Angular distance from the BooIII center derived by \citet{Grillmair2009}.}
\end{center}
\end{table}

\end{document}